\documentclass[twocolumn,aps,prl,superscriptaddress,amsmath,amssymb,superscriptaddress]{revtex4-2}

\usepackage{graphicx}
\usepackage{float}
\usepackage{natbib}
\usepackage{amsmath}
\graphicspath{ {./images/} }
\usepackage{xparse}
\usepackage{siunitx}
\usepackage{hyperref}
\usepackage{lipsum}
\usepackage{braket}
\usepackage{nicefrac, xfrac}
\newcommand{\mbeq}{\overset{!}{=}}

\begin{document}

\title{A single-photon emitter coupled to a phononic-crystal resonator in the resolved-sideband regime}
\author{Clemens Spinnler}
\email{c.spinnler@unibas.ch}

\affiliation{Department of Physics, University of Basel, Klingelbergstrasse 82, CH-4056 Basel, Switzerland}

\author{Giang N. Nguyen}
\affiliation{Department of Physics, University of Basel, Klingelbergstrasse 82, CH-4056 Basel, Switzerland}

\author{Ying Wang}
\affiliation{Center for Hybrid Quantum Networks (Hy-Q), The Niels Bohr Institute, University of Copenhagen, DK-2100 Copenhagen Ø, Denmark}

\author{Liang Zhai}
\altaffiliation[Current address: ]{Pritzker School of Molecular Engineering, University of Chicago, Chicago, IL 60637, USA}
\affiliation{Department of Physics, University of Basel, Klingelbergstrasse 82, CH-4056 Basel, Switzerland}

\author{Alisa Javadi}
\altaffiliation[Current address: ]{School of Electrical and Computer Engineering, University of Oklahoma, Norman, OK 73019, USA}
\affiliation{Department of Physics, University of Basel, Klingelbergstrasse 82, CH-4056 Basel, Switzerland}

\author{Marcel Erbe}
\affiliation{Department of Physics, University of Basel, Klingelbergstrasse 82, CH-4056 Basel, Switzerland}

\author{Sven Scholz}
\affiliation{Lehrstuhl für Angewandte Festkörperphysik, Ruhr-Universität Bochum, DE-44780 Bochum, Germany}

\author{Andreas D. Wieck}
\affiliation{Lehrstuhl für Angewandte Festkörperphysik, Ruhr-Universität Bochum, DE-44780 Bochum, Germany}

\author{Arne Ludwig}
\affiliation{Lehrstuhl für Angewandte Festkörperphysik, Ruhr-Universität Bochum, DE-44780 Bochum, Germany}

\author{Peter Lodahl}
\affiliation{Center for Hybrid Quantum Networks (Hy-Q), The Niels Bohr Institute, University of Copenhagen, DK-2100 Copenhagen Ø, Denmark}

\author{Leonardo Midolo}
\affiliation{Center for Hybrid Quantum Networks (Hy-Q), The Niels Bohr Institute, University of Copenhagen, DK-2100 Copenhagen Ø, Denmark}

\author{Richard J. Warburton}
\affiliation{Department of Physics, University of Basel, Klingelbergstrasse 82, CH-4056 Basel, Switzerland}

\date{\today} 

\begin{abstract}
A promising route towards the heralded creation and annihilation of single-phonons is to couple a single-photon emitter to a mechanical resonator. The challenge lies in reaching the resolved-sideband regime with a large coupling rate and a high mechanical quality factor. We achieve all of this by coupling self-assembled InAs quantum dots to a small-mode-volume phononic-crystal resonator with mechanical frequency $\Omega_\mathrm{m}/2\pi = $ \qty{1.466}{\GHz} and quality factor $Q_\mathrm{m} = 2.1\times10^3$. Thanks to the high coupling rate of $g_\mathrm{ep}/2\pi = $\qty{2.9}{\MHz}, and by exploiting a matching condition between the effective Rabi and mechanical frequencies, we are able to observe the interaction between the two systems. Our results represent a major step towards quantum control of the mechanical resonator via a single-photon emitter.

\end{abstract}

\maketitle
Coupling a quantum system to a mechanical resonator is of both fundamental and technological interest. A highly studied area is the use of photons to control the state of a mechanical system~\cite{Aspelmeyer2014}. To enhance the interaction, cavities have been introduced in the form of two mirrors facing each other~\cite{Thompson2008}, ring resonators~\cite{Verhagen2012}, photonic crystals~\cite{Chan2011}, and microwave resonators~\cite{Teufel2011}. Various quantum systems can be coupled to mechanical resonators, for instance superconducting qubits~\cite{Chu2017}, atomic ensembles~\cite{Karg2020}, and solid-state emitters~\cite{DeCrescent2022}.\\
\indent For phononic quantum technologies~\cite{Barzanjeh2022}, it becomes necessary to create one phonon at a time~\cite{Riedinger2016}. Using optical cavities, this is achieved with a highly attenuated laser pulse; the phonon creation probability is much below one~\cite{Riedinger2016}. An alternative approach is to use a single-photon emitter. In combination with phononic and photonic waveguides, it was proposed that such a coupled system can serve as a deterministic source of single-phonons~\cite{Sollner2016}.\\
\indent To date, different kinds of single-photon emitters have been coupled to mechanical resonators: colour centres~\cite{,Kenneth2016,Cady_2019}, rare-earth ions~\cite{Ohta2021}, 2D-materials~\cite{ Patel2022}, and quantum dots~\cite{Carter2018,Carter2019,Yuan2019,Vogele2020,Montinaro2014,Descamps2023,Yeo2014, Yeo2016, Assis2017, Kettler2021, Finazzer2023,Carter2017,Nysten2020,Imany2022,DeCrescent2022}. So far, the sensitivity to mechanical motion is low and the interaction between the two systems was probed by driving the mechanical resonator, i.e., by adding phonons. However, to explore single-phonon applications, it is crucial to operate in the few-phonon regime, ultimately in the quantum ground state~\cite{oConnell2010}.\\
\indent Semiconductor quantum dots (QDs) are excellent emitters of high-quality single-photons~\cite{Senellart_2017,Lodahl2022,Zhai2020,Zhai2022}. The upper level, the exciton, couples to lattice vibrations via a strain-induced deformation potential~\cite{WilsonR2004,Munsch2017,Yeo2014}. In the resolved-sideband regime, the angular frequency of the mechanical resonance, $\Omega_\mathrm{m}$, must be larger than the radiative decay rate of the QD exciton, $\Gamma_\mathrm{R}$. In addition, for optical driving of the phonon sidebands, it is important that $\Omega_\mathrm{m}$ is also larger than the inhomogeneously-broadened linewidth, $\Gamma_\mathrm{inh}$~\cite{spinnler_open_2023}. Satisfying these two conditions is challenging for many reasons. First, the mechanical frequencies need to be in the gigahertz regime, where mechanical losses tend to be high~\cite{spinnler_open_2023}. Second, obtaining narrow optical linewidths (i.e., small inhomogeneous broadening) on QDs embedded in mechanical resonators can be difficult~\cite{Munsch2017}. Third, for high exciton-phonon coupling rates, the mechanical resonator size needs to be small such that the implementation of a phononic shield becomes necessary~\cite{Safavi2010}.\\
\indent In this work, we face the aforementioned challenges by coupling self-assembled InAs QDs \cite{Loebl2019} to a phononic-crystal resonator (PnCR) with $\Omega_\mathrm{m}/2\pi = $ \qty{1.466}{\GHz} and $Q_\mathrm{m} = 2.1\times10^3$. 
We achieve narrow optical linewidths and hence a high sensitivity of the QD's optical response to mechanics. For the first time, we detect the interaction of a quantum emitter and a mechanical resonator in the few-phonon regime with $\langle n_\mathrm{m}\rangle = 58$ (thermal motion at \qty{4.2}{\K}). We drive the QD's optical transition resonantly and analyse the emitted photons using quantum optics techniques, specifically a measurement of the auto-correlation function. 
\begin{figure}[ht]
\includegraphics{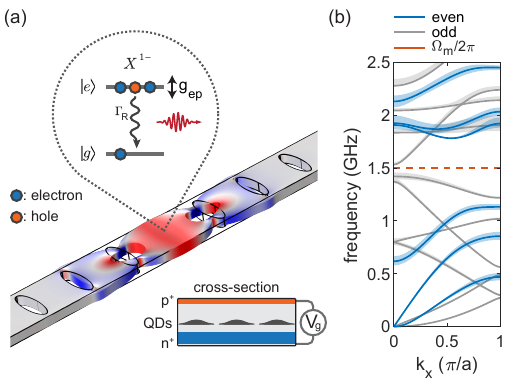}
\caption{\label{fig:1} \textbf{Finite-element simulations of a QD coupled to a mechanical resonator.} \textbf{(a)} Phononic-crystal resonator hosting a semiconductor diode-structure for QD charge control (see cross-section). The QD emits single photons and its excited state is dispersively coupled to the mechanical motion via a deformation potential (example given for the negative trion $X^{1-}$). The resonator consists of a well-isolated mechanical mode at $\Omega_\mathrm{m}/2\pi\approx$ \qty{1.5}{\GHz} tightly confined by the surrounding phononic shield. \textbf{(b)} Band diagram of the phononic shield. The width of the bandgap is \qty{0.65}{\GHz} and \qty{0.11}{\GHz} for even and odd modes in terms of the $z$-symmetry, respectively. Shaded areas represent band broadening upon varying the air-hole parameters by $\pm 20~\mathrm{nm}$.}
\end{figure}
We show that the influence of the mechanical resonance on the autocorrelation satisfies a resonance condition: the effective Rabi-frequency should match the mechancial resonance-frequency. We extract a state-of-the-art exciton-phonon coupling rate of $g_\mathrm{ep}/2\pi = $ \qty{2.9}{\MHz}. \\ 
\indent The mechanical resonator is etched into a \qty{180}{\nm} GaAs membrane. The membrane hosts InAs QDs; the QDs are embedded in a p-i-n diode, see inset to Fig.~\ref{fig:1}(a)~\cite{Midolo2015,Lobl2017,Ludwig2018}. Figure~\ref{fig:1}(a) and (b) show finite-element simulations of the PnCR. The mechanical in-plane breathing mode is tightly confined by the phononic bandgap structure~\cite{Safavi2010} and shows a highly homogeneous strain profile in the centre of the resonator (see also Supplement). The PnCR consists of seven holes (phononic-shield unit cells) etched into the membrane~\cite{Midolo2015}, see Fig.~\ref{fig:2}(a). Due to the small beam cross-section ($960 \times$\qty{180}{\nm^2}), the breathing mode has a mode volume as small as $4\times10^{-3}\lambda^3$ (with $\lambda=$ \qty{3.25}{\um}). As a result, our simulations predict a low effective mass $m_\mathrm{eff} = 7.4 \times10^{-16}~\mathrm{kg}$, a large zero-point motion $x_\mathrm{zpf} = 2.7 \times10^{-15}~\mathrm{m}$, and a high exciton-phonon coupling rate $g_\mathrm{ep}^\mathrm{sim}/2\pi= $\qty{3.2}{\MHz} (see Supplement). The thermal phonon occupation of the breathing mode is $\langle n_\mathrm{m}\rangle = 58$ (at \qty{4.2}{\K}). A crucial point of our design is that we also optimise the resonator geometry in terms of the optical outcoupling of the emitted photons (see Ref.~\cite{spinnler_open_2023}).\\
\indent We select a QD with a potentially large exciton-phonon coupling rate from a photoluminescence map~\cite{Loebl2019b,spinnler_open_2023}.   
Fig.~\ref{fig:2}(b) shows several bright QDs located around the centre of the beam. Switching to resonant excitation of a single QD, a plateau map of the negative trion $X^{1-}$ is recorded, see Fig.~\ref{fig:2}(c). Upon changing the applied gate voltage, $V_\mathrm{g}$, the emission frequency is tuned (via the dc Stark effect) over a frequency range of more than 50 times the QD's linewidth. The lifetime of the excited state is extracted from a pulsed measurement $\tau_\mathrm{R} = $ \qty{1.18}{\ns}, which corresponds to a radiative decay rate of $\Gamma_\mathrm{R} = 2\pi\times135~\mathrm{MHz}$ (see Supplement). On account of the diode structure and optimised nano-fabrication, we obtain narrow optical linewidths, here, a factor of four above the transform limit $\Gamma_\mathrm{inh} = 4\Gamma_\mathrm{R}$, see Fig.~\ref{fig:2}(d). \\
\begin{figure}[b]
\includegraphics{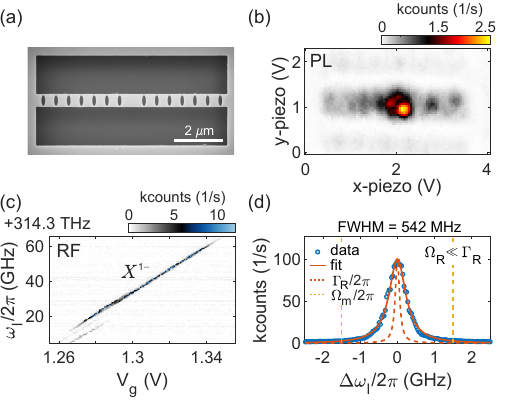}
\caption{\label{fig:2} \textbf{QD characterisation.} \textbf{(a)} Scanning electron-microscope image of the fabricated device with seven shield-elements on each side. \textbf{(b)} Photoluminescence map showing several QDs in the centre of the phononic-crystal beam. \textbf{(c)} Resonance fluorescence charge-plateau scan of the negative trion ($X^{1-}$) of a QD close to the resonator's centre. \textbf{(d)} Low-power frequency scan to determine the inhomogeneously-broadened linewidth, $\Gamma_\mathrm{inh}/2\pi = $ \qty{550}{\MHz}. The data are fitted to a Lorentzian. The transform limit and the sideband position ($\Delta\omega_\mathrm{l} = \pm\Omega_\mathrm{m}$) are shown. }
\end{figure}
\indent The only way to detect the specific mechanical mode is via the QD. Thus, it is crucial to consider how the exciton-phonon coupling can be detected, in particular the coupling to mechanical oscillations driven only by thermal noise (i.e., the Brownian motion). In the unresolved-sideband regime, mechanical noise results in fluctuations in the intensity of the resonance fluorescence~\cite{Munsch2017}. The relative sensitivity is proportional to the square of the normalised derivative of the resonance fluorescence counts (with respect to laser detuning)~\cite{Munsch2017}, see Fig.~\ref{fig:3}(b). The highest sensitivity is obtained at low excitation power and laser detunings corresponding to half the optical linewidth. In the resolved sideband regime, this technique fails. The mechanical resonator oscillates back-and-forth several times during the exciton lifetime, resulting in acoustic sidebands at $\Delta\omega_\mathrm{l} = \pm\Omega_\mathrm{m}$~\cite{spinnler_open_2023}, such that the resonance fluorescence intensity becomes insensitive to the mechanical motion.\\
\indent To find an alternative detection scheme, we simulate a driven two-level system. The mechanical coupling is introduced via a frequency shift of the upper level, here the trion. The model is semi-classical: the two-level system is treated quantum mechanically; both the laser drive and the mechanics are treated classically. 
\begin{figure*}[ht]
    \centering
    \includegraphics[width=1\textwidth]{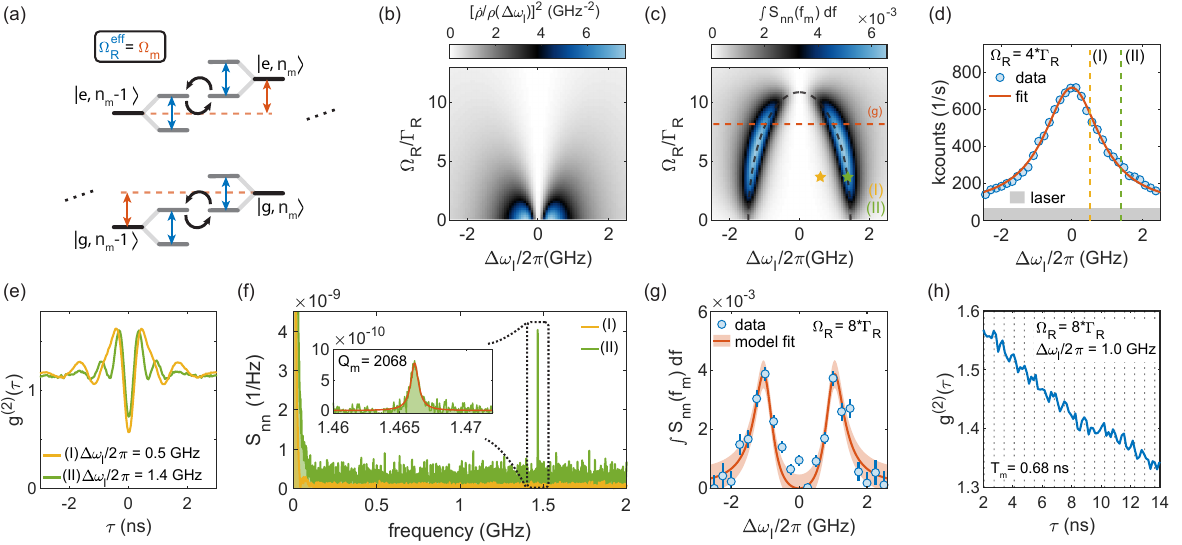}
    \caption{\textbf{Exciton-phonon coupling strength.} \textbf{(a)} Phonon-energy-ladder diagram for a driven two-level system. The excitation laser dresses the two-level system. The interaction between the two systems is strongest when the dressed-state splitting (effective Rabi frequency, $\Omega_\mathrm{R}^\mathrm{eff}/2\pi$) is equal to the sideband separation (mechanical frequency, $\Omega_\mathrm{m}/2\pi$). \textbf{(b)} Simulated mechanical-noise sensitivity in the unresolved-sideband regime, given by the derivative of the QD counts. \textbf{(c)} Simulated mechanical-noise sensitivity of the current device. $\int S_\mathrm{nn}(f_\mathrm{m})df$ is the integrated noise power of the mechanical resonace. The condition of $\Omega_\mathrm{R}^\mathrm{eff}=\Omega_\mathrm{m}$ is highlighted by the dashed black line. \textbf{(d)} QD linewidth at elevated excitation power ($\Omega_\mathrm{R} = 4\Gamma_\mathrm{R}$) as used in the thermal-motion measurement. \textbf{(e)} Autocorrelation measurements at two different detunings (see dashed lines in (d)). The larger detuning corresponds to $\Omega_\mathrm{R}^\mathrm{eff}=\Omega_\mathrm{m}$. \textbf{(f)} Noise-power spectra obtained from (e) reveal the mechanical peak only for optimal effective Rabi frequency. \textbf{(g)} Integrated mechanical noise in dependence on the laser detuning. An exciton-phonon coupling rate of $g_\mathrm{ep}/2\pi =$ \qty{2.9}{\MHz} is extracted from a model-fit to the data. Data and fit errors are given by one standard deviation. \textbf{(h)} Zoom-in of a $14-\mathrm{hours}$ autocorrelation measurement, showing mechanical oscillations. Dashed lines show the expected noise peaks, spaced by $T_\mathrm{m}=2\pi/\Omega_\mathrm{m} = $ \qty{0.68}{\ns}. }
    \label{fig:3}
\end{figure*}
We assume that the mechanical phase is static with respect to the time dynamics of the quantum emitter~\cite{Weiss2021,Wigger2021}. The master equations are complex and we therefore solve them numerically. Specifically, we simulate the sensitivity of the resonance fluorescence to the mechanical motion, $\int S_\mathrm{nn}(f_\mathrm{m})df$ ($S_\mathrm{nn}$ being the noise-power spectrum), as a function of laser detuning and Rabi frequency, Fig.~\ref{fig:3}(c). As expected, the sensitivity is zero at small detunings and at small Rabi couplings. Strikingly, a mechanical signal is obtained for specific combinations of excitation powers and laser detunings.\\
\indent A qualitative understanding of the sensitivity resonances in Fig.~\ref{fig:3}(c) can be found in the dressed-state picture. Figure~\ref{fig:3}(a) shows the phonon-ladder diagram where neighbouring ground and excited states are separated by the phonon energy $\hbar\Omega_\mathrm{m}$ (orange arrows). Vertical transitions correspond to transitions which conserve the phonon number of the resonator (elastic scattering) and diagonal transitions change the phonon population by $\pm1$ (Stokes/anti-Stokes scattering). The optical excitation dresses the energy levels of the quantum emitter and splits them by an energy $\hbar \Omega_\mathrm{R}^\mathrm{eff} = \hbar \sqrt{\Omega_\mathrm{R}^2+\Delta\omega_\mathrm{l}^2}$ (blue arrows), where $\Omega_\mathrm{R}$ is the Rabi coupling and $\Delta\omega_\mathrm{l}/2\pi$ is the laser detuning. The sensitivity of the QD to the mechanical resonator is strongest whenever a combination of $\Omega_\mathrm{R}$ and $\Delta\omega_\mathrm{l}$ is chosen such that $\Omega_\mathrm{R}^\mathrm{eff} = \Omega_\mathrm{m}$~\cite{Rabl_cooling_2010}. In the dressed-state picture this means that states involving different phonon numbers are degenerate and mechanical quanta can be exchanged without the cost of energy (black arrows).\\
\indent Creating a time modulation in the RF intensity can be understood qualitatively in the sideband picture. The red and blue sidebands oscillate with $\Omega_\mathrm{m}/2\pi$, however, out-of-phase~\cite{Wigger2021}. On resonance, the modulations cancel out. By red (blue) detuning the probe laser, an imbalance between Stokes and anti-Stokes sidebands is created. This imprints a time modulation in the QD signal and also leads to cooling (heating) of the mechanical resonator.\\ 
\indent According to the simulations, the optimal laser detuning for the highest sensitivity to mechanical motion should satisfy $\Delta\omega_\mathrm{l}=\sqrt{\Omega_\mathrm{m}^2-\Omega_\mathrm{R}^2}$ (dashed black line in Fig.~\ref{fig:3}(c)). At low excitation powers ($\Omega_\mathrm{R} \ll \Gamma_\mathrm{m}$) this means detuning the laser to the blue or red sideband. At high excitation powers ($\Omega_\mathrm{R} > \Gamma_\mathrm{m}$), the laser detuning needs to be smaller than the mechanical frequency. Above $\Omega_\mathrm{R}>\Omega_\mathrm{m}$ ($\Omega_\mathrm{m} = 10.8\Gamma_\mathrm{R}$ here), the resonance condition can no longer be met and the sensitivity is therefore small. The signal-to-noise ratio increases with higher QD counts, with a maximum around $\Omega_\mathrm{R} \approx 8\Gamma_\mathrm{R}$, (see Supplement).\\
\indent The simulations motivate the choice of $\Omega_\mathrm{R}^\mathrm{eff}$ in the experiment. We perform two individual measurements with parameter sets (Rabi frequency and laser detuning) (I) $\Omega_\mathrm{R}^\mathrm{eff} < \Omega_\mathrm{m}$ and (II) $\Omega_\mathrm{R}^\mathrm{eff} = \Omega_\mathrm{m}$. Here, $\Omega_\mathrm{m}$ is the mechanical frequency expected from the simulations. The measurements are performed at $\Omega_\mathrm{R} = 4\Gamma_\mathrm{R}$, where the linewidth is just slightly power broadened. $\Omega_\mathrm{R}$ is calibrated with a power series measurement (see Supplement). Figure~\ref{fig:3}(d) shows the associated linewidth scan. Due to the high optical power, the background level (unsuppressed laser) increases. For (I) the laser detuning corresponds to $\Delta\omega_\mathrm{l}/2\pi =$ \qty{0.5}{\GHz} (yellow line) and for (II) $\Delta\omega_\mathrm{l}/2\pi =$ \qty{1.4}{\GHz} (green line). \\
\indent To probe the QD-mechanical interaction we record $8~\mathrm{hours}$ of autocorrelation data while repeatedly suppressing the laser and correcting for a small spectral drift (see Supplement). Figure~\ref{fig:3}(e) shows the autocorrelation measurements for (I) and (II). Close to zero delay, Rabi oscillations of $\Omega_\mathrm{R}^\mathrm{eff}/2\pi$ are visible, with frequencies matching those expected from the calibration. The dip at $\tau=0$ (a measure of the single-photon purity) reaches values far from zero on account of the increased laser background (see Fig.~\ref{fig:3}(d) and Supplement). The noise-power spectrum, $S_\mathrm{nn}$, is obtained from the autocorrelation measurement via Fourier transform (see Supplement)~\cite{Munsch2017}. Figure~\ref{fig:3}(f) shows $S_\mathrm{nn}$ for configurations (I) and (II) obtained from an autocorrelation with a maximum time delay of $\tau_\mathrm{max}=$ \qty{500}{\ns} and time binning of $\tau_\mathrm{bin}=$ \qty{50}{\ps}. In case (II) (green spectra), a prominent peak is visible at a frequency of \qty{1.466}{\GHz}. In case (I) (yellow spectra), no peaks are visible, although the photon-click rate on the detectors is much higher and the noise floor is much lower than for (II). We identify the peak as the mechanical resonance on two grounds. First, $S_\mathrm{nn}$ follows the dependence on $\Omega_\mathrm{R}^\mathrm{eff}$ as predicted by the simulations. Second, the frequency at the peak, \qty{1.466}{\GHz} lies very close to the mechanical resonance as predicted by the simulations, \qty{1.496}{\GHz}. Thus, the coupled hybrid system is in the resolved-sideband regime with $\Omega_\mathrm{m} \approx 11\Gamma_\mathrm{R}$ but more importantly $\Omega_\mathrm{m}\approx2.7\Gamma_\mathrm{inh}$.\\
\indent The inset to Fig.~\ref{fig:3}(f) displays the mechanical noise-peak with high resolution (FFT of autocorrelation with $\tau_\mathrm{max}=$ \qty{8}{\us} and $\tau_\mathrm{bin}=$ \qty{50}{\ps}). A mechanical quality factor as high as $Q_\mathrm{m}=2.1\times10^3$ is extracted via a Lorentzian fit. This justifies the assumption that the mechanical damping rate is much less than than the QD's radiative decay rate. \\
\begin{figure}[t]
\includegraphics{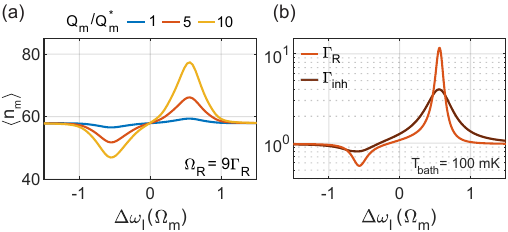}
\caption{\label{fig:4} \textbf{Simulation of mechanical backaction.} \textbf{(a)} Expectation value of the phonon population of the mechanical resonator with increasing mechanical quality factor at \qty{4.2}{\K}. $Q_\mathrm{m}^*$ corresponds to the current device parameter of $2.1\times10^3$. \textbf{(b)} Phonon occupation of the resonator with $Q_\mathrm{m}=10^4$ and $g_\mathrm{ep}/2\pi=2\times3.2~\mathrm{MHz}$ on starting at millikelvin temperatures, with (dark line) and without (bright line) inhomogenous broadening. For details on the simulations see Supplement. }
\end{figure} 
\indent To estimate the exciton-phonon coupling rate, ${g_\mathrm{ep}=\left(\delta\omega_\mathrm{QD}/\delta x\right)\cdot x_\mathrm{zpf}}$, we perform the noise-power measurement upon sweeping the laser detuning (at fixed laser power). We chose $\Omega_\mathrm{R} = 8\Gamma_\mathrm{R}$, for which the highest interaction between the two systems is expected at $\Delta\omega_\mathrm{l}/2\pi= $ \qty{1.0}{\GHz}. Figure~\ref{fig:3}(g) shows the integrated noise power versus laser detuning. For each data point, $1~\mathrm{hour}$ of autocorrelation data are recorded. Via a model fit to the data, an exciton-phonon coupling strength of $g_\mathrm{ep}/2\pi =$ \qty{2.9}{\MHz} is extracted. This agrees very well with the simulated value (\qty{3.2}{\MHz}). Finally, we repeat the autocorrelation measurement with optimised parameters for the highest signal-to-noise ratio, see Fig.~\ref{fig:3}(h). The oscillations due to the interaction with the mechanical resonator now become visible, even without a Fourier transform.  \\
\indent A prominent feature of the noise spectra, Fig.~\ref{fig:3}(f), is that only one peak is observed. Furthermore, both the mechanical frequency and the coupling $g_\mathrm{ep}/2\pi$ are very close to the simulated values. These observations show that the phononic resonator operates as per design: modes in the gap are both decoupled from the environment via the phononic shield and highly confined to the central island. The many mechanical modes lying outside the bandgap are highly damped and are therefore not observed. Considering that the phononic shield works well, it is likely that the mechanism limiting the mechanical quality factor is related to damping mechanisms within the material. Notably, GaAs gigahertz mechanical resonators with mechanical quality factors exceeding $10^4$ have been achieved~\cite{Forsch2020,DeCrescent2022}. This suggests that the present mechanical quality factor is limited by surface losses and/or losses within the doped regions. One approach to reduce these losses could be surface passivation~\cite{Kuruma2020,Najer2021,Chan2012thesis}. Moreover, the quality factor could also be enhanced via soft clamping and further strain engineering~\cite{Tsaturyan2017,Ghadimi2018}.\\ 
\indent The figure of merit for the mechanical signal strength is the thermal coupling rate normalised to the mechanical angular frequency, $g^\mathrm{th}_\mathrm{ep}/\Omega_\mathrm{m}$ with $g^\mathrm{th}_\mathrm{ep} = g_\mathrm{ep}\sqrt{2\langle n_\mathrm{m}\rangle}$ (see Supplement). To enhance the signal strength, it is important to improve $g_\mathrm{ep}$ further without increasing the mechanical frequency. By choosing a two-dimensional resonator design hosting a degenerate breathing mode, the coupling could be enhanced about two-fold.\\
\indent For single-phonon experiments it is crucial to reach below $\langle n_\mathrm{m}\rangle=1$. To estimate the backaction on the mechanical resonator, we perform additional simulations in which the mechanical resonator is described in a fully quantum way with decay to a thermal bath. Also here, the strongest interaction is found when $\Omega_\mathrm{R}^\mathrm{eff}=\Omega_\mathrm{m}$. Currently, the change in phonon number that we achieve with red-detuned driving is small (below two). The cooling performance can be improved by either increasing the coupling strength, reducing the thermalisation rate to the bath, or reducing the inhomogeneous broadening. At \qty{4.2}{\K}, measurable cooling can be observed with a five-fold increased mechanical quality factor ($Q_\mathrm{m}=10^4$), however, the lowest phonon number is still much above one, see Fig.~\ref{fig:4}(a). An alternative approach is to start at a bath temperature of $100~\mathrm{mK}$, where $\langle n_\mathrm{m}\rangle=0.98$. With a five-fold enhanced $Q_\mathrm{m}$, a two-fold increased $g_\mathrm{ep}$, and a transform-limited linewidth~\cite{Pedersen_2020}, the lowest phonon occupation that can be reached is $\langle n_\mathrm{m}\rangle=0.56$, a suitable starting point for single-phonon experiments, see Fig.~\ref{fig:4}(b).\\
\indent In summary, we present a QD coupled to a phononic-crystal resonator in the resolved-sideband regime. Upon resonant excitation, the QD has a narrow optical linewidth and a high mechanical sensitivity. We demonstrate that the interaction between the two systems is strongest when matching the eigenfrequency of the dressed QD with the mechanical frequency: $ \Omega_\mathrm{R}^\mathrm{eff} =  \Omega_\mathrm{m}$. Satisfying this condition allows us to observe even the thermally-driven mechanical oscillations (at \qty{4.2}{\K}) and to determine precisely the mechanical resonator's frequency. This represents a crucial way to characterise hybrid mechanical devices with single-photon emitters. Via the noise power-spectrum, we determine the exciton-phonon coupling, $g_\mathrm{ep}/2\pi =$ \qty{2.9}{\MHz} (an unprecedentedly high value) and the mechanical $Q$-factor, $2.1\times10^3$. The low mechanical damping rate results in a high $Q_\mathrm{m}\cdot f_\mathrm{m}$ product of $3.0 \times 10^{12}~\mathrm{Hz}$, which allows for $34$ mechanical oscillations before coherence is lost. We show that with slightly improved device parameters, cooling and heating of the mechanical resonator via optical excitation of the QD can be observed. Thanks to the gigahertz mechanical frequency, cooling to millikelvin temperatures is sufficient to reach $\langle n_\mathrm{m}\rangle<1$. In this regime, several QDs can be coupled to the same mechanical resonator facilitating a coherent exciton-exciton coupling.

\section*{Acknoledgements}
We thank Martino Poggio, Tomasz A.\ Jakubczyk, Yannik Laurent Fontana, Hinrich Mattiat, Thibaud Ruelle, and Guillaume Bertel for stimulating discussions.\\
\indent The work was supported by SNF Project 200020\_204069. GNN, ME, LZ, and AJ received funding from the European Union’s Horizon 2020 Research and Innovation Programme under the Marie Sk\l{}odowska-Curie grant agreement No.\ 861097 (QUDOT-TECH), No.\ 721394 (4PHOTON),  and No.\ 840453 (HiFig), respectively. YW, PL and LM acknowledge financial support from Danmarks Grundforskningsfond (DNRF 139, Hy-Q Center for Hybrid Quantum Networks). LM acknowledges the European Research Council (ERC) under the European Union’s Horizon 2020 research and innovation programme (Grant Agreement No. 949043, NANOMEQ). SS, ADW and AL acknowledge financial support from the grants DFH/UFA CDFA05-06, DFG TRR160, DFG project 383065199, and BMBF Q.Link.X 16KIS0867. 

\section*{Competing Interests}
The authors declare no competing interests.

\section*{Data Availability}
The data that supports this work is available from the corresponding author upon reasonable request.

\section*{Code Availability}
The code used for this work is available from the corresponding author upon reasonable request.


%

\onecolumngrid
\newpage

\section*{Supplementary Information: A single-photon emitter coupled to a phononic-crystal resonator in the resolved-sideband regime}
\setcounter{section}{0}

\section*{I. Finite-element simulations}
The mechanical resonator is simulated using Comsol Multiphysics. We perform eigenmode studies for a predefined mesh, which is optimised via a convergence test. Either fixed or lossy boundary conditions are applied to the edges of the beam. The phononic shield and the mechanical resonator are optimised to support a single mechanical in-plane breathing mode within the phononic bandgap. Figure~\ref{fig:1s}(a) shows the top view of the phononic shield's unit cell. The dimensions of the etched air-hole are: $a = $ \qty{970}{\nm} (lattice constant), $w = $ \qty{960}{\nm} (beam width), $h = $ \qty{180}{\nm} (beam thickness), $l_\mathrm{e} = $ \qty{775}{\nm} (air-hole length), and $w_\mathrm{e} = $ \qty{270}{\nm} (air-hole width). For quantum-dot (QD) charge control, we need to ensure that the gate layers show good conductance from the substrate through the phononic shield to the resonator. Therefore we try to keep the thinnest connection, between the resonator and substrate, as wide as possible, which is $(w-l_\mathrm{e})/2 = $ \qty{92.5}{\nm}. For the current design, we observe about \qty{100}{\MHz} of additional linewidth broadening, for more information see Ref.~\cite{spinnler_open_2023}.\\
\indent The phononic band diagram of the unit-cell is studied by applying floquet boundary conditions in the $x$-direction. Eigenmode studies are performed upon sweeping $k_\mathrm{x}$ from $0$ to $\pi/a$, where $a$ is the lattice constant. For even and odd modes we apply symmetric and asymmetric conditions at $z=0$, respectively. The band diagram is shown in Fig.~1(b) in the main text. We observe a complete bandgap of \qty{0.11}{\GHz}. We repeat the study and change the air-hole parameters by $\pm 20~\mathrm{nm}$. Although this leads to a broadening of the bands, the width of the phononic gap is not much affected (especially for even modes). This suggests that the phononic shield is relatively robust against small deviations from the original design. From the band diagram study, we can also extract the density of states (DOS), shown in Fig.~\ref{fig:1s}(b). We sum the obtained eigenmodes over a specific frequency bandwidth and normalise the states to the unit-cell length~\cite{Florez2022}: 
\begin{equation}
    \mathrm{DOS}=\frac{n_{\Delta f}}{\Delta  f\cdot n_\mathrm{k} \cdot a},
\end{equation}
where $\Delta  f$ is the frequency bandwidth, $n_{\Delta f}$ is the number of modes in $\Delta  f$, $n_\mathrm{k}$ is the number of $k$-points in our study, and $a$ is the unit-cell length. \\
\indent To estimate how many phononic-shield elements are needed, such that the resonator's damping is not limited by clamping losses, we perform a mechanical-quality-factor study in dependence on the number of shield unit cells, see Fig.~\ref{fig:1s}(c). The study is performed by applying a lossy boundary condition at the two beam ends. Since no phonon loss within the material is added, this only simulates clamping losses. We select seven air holes (yellow line) on each side of the resonator.\\
\indent To finalise our design, we perform an optical far-field study of an emitter placed in the centre of the mechanical resonator and we optimise the beam width and resonator defect length ($x$-direction) for radiation to the top (collection lens). For a width of \qty{960}{\nm}, the two sidewalls reflect the field within the beam such that there is constructive interference, increasing the radiation of the field to the top~\cite{spinnler_open_2023}.\\
\indent  To estimate the exciton-phonon coupling rate, we perform thermomechanical calibration~\cite{Hauer2013}. We describe the system by a linear harmonic oscillator via the displacement function: 
\begin{equation}
u(r,t) =x(t)\lvert u(r)\rvert,
\end{equation}
where $\lvert u(r)\rvert = \frac{u(r)}{\mathrm{max}(\lvert u(r) \rvert)}$ describes the normalised mode profile~\cite{Hauer2013} obtained from COMSOL, and $x(t)$ describes the time dependence of the periodic motion. The equation of motion is: 
\begin{figure*}[ht]
    \centering
    \includegraphics[width=0.9\textwidth]{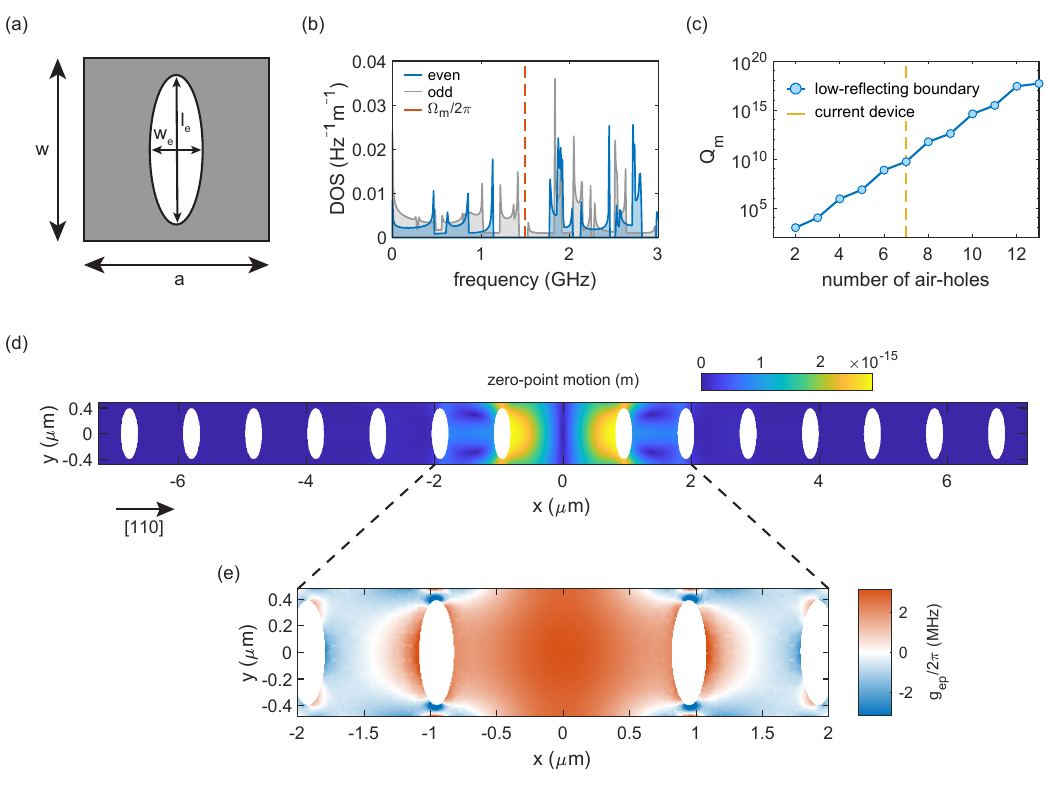}
\caption{\label{fig:1s} \textbf{Finite-element simulations.} \textbf{(a)} Top view of the phononic-shield unit cell which consists of an elliptical air-hole: $\mathrm{a} = $ \qty{970}{\nm} (lattice constant), $\mathrm{w} = $ \qty{960}{\nm}, $\mathrm{h} = $ \qty{180}{\nm} (beam thickness), $\mathrm{l_e} = $ \qty{775}{\nm}, and $\mathrm{w_e} = $ \qty{270}{\nm}. \textbf{(b)} Mechanical density of states (DOS) for even (in-plane/symmetric) and odd (out-of-plane/asymmetric) modes. The mechanical frequency is designed to lie in the centre of the complete bandgap. The simulation is performed with $n_\mathrm{k}=2000$ and $\Delta f = $ \qty{8}{\MHz}. \textbf{(c)} Mechanical quality factor with increasing number of shield unit cells. For the device presented in the main paper, seven unit cells were chosen. \textbf{(d)} Thermal displacement of an eigenmode study obtained via thermomechanical calibration. The major displacement of the in-plane breathing mode is along the beam axis. \textbf{(e)} Exciton-phonon coupling rate based on deformation potential coupling, evaluated in the centre of the membrane (QD layer). A coupling rate of $g_\mathrm{ep}/2\pi = $ \qty{3.2}{\MHz} is found in the centre of the resonator. }
\end{figure*}
\begin{equation}
m_\mathrm{eff}\frac{dx^2(t)}{dt^2}+m_\mathrm{eff}\Gamma_\mathrm{m}\frac{dx(t)}{dt}+m_\mathrm{eff}\Omega_\mathrm{m}^2
x(t) = F(t),
\end{equation}
where $\Omega_\mathrm{m}/2\pi$ is the mechanical frequency, $k = m_\mathrm{eff}\Omega_\mathrm{m}^2$ is the spring constant, and $\Gamma_\mathrm{m}$ is the energy dissipation rate which relates to the mechanical quality by $Q_\mathrm{m} = \Omega_\mathrm{m}/\Gamma_\mathrm{m}$. The effective mass and zero-point-motion are obtained with: 
\begin{equation}
    m_{\mathrm{eff}} = \int \rho  \left( \frac{u(r)^2}{\mathrm{max}(\lvert u(r) \rvert)^2} \right) \,dV
\end{equation}

\begin{equation}
    x_{\mathrm{zpf}} = \sqrt{\frac{\hbar}{2m_{\mathrm{eff}}\Omega_\mathrm{m}} },
\end{equation}
where $\rho$ is the material density of GaAs, and $\hbar$ is the reduced Planck constant. Figure~\ref{fig:1s}(d) shows the displacement, $x_\mathrm{zpf}\cdot\lvert u(r)\rvert$, of the in-plane breathing mode. The thermal displacement (Brownian-motion) is then given by the equipartition theorem~\cite{Yeo2014}:
\begin{equation}
    x_{\mathrm{th}} = x_{\mathrm{zpf}}\sqrt{\frac{2k_\mathrm{B}T}{\hbar \Omega_\mathrm{m}}},
\end{equation}
where $T$ is the phonon-bath temperature, and $k_\mathrm{B}$ is the Boltzmann constant. 
\begin{figure*}[ht]
    \centering
    \includegraphics[width=0.95\textwidth]{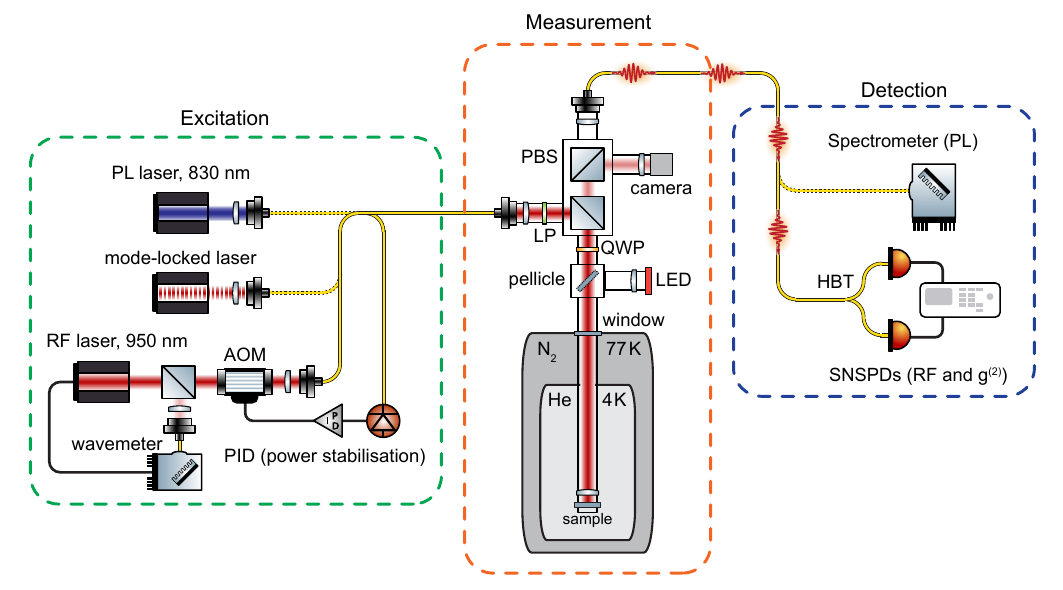}
\caption{\label{fig:5s} \textbf{Optical setup including all relevant hardware for excitation, measurement, and detection.} A double-pass acousto-optic modulator setup is used to stabilise the RF laser power. The excitation laser is fibre-coupled and sent to the dark-field microscope. With two PBS, an LP, and a QWP, the reflected laser light is suppressed. The collected QD single-photons are also fibre-coupled and analysed using SNSPDs. PL: photoluminescence, RF: resonance fluorescence, PBS: polarising beam splitter, LP: linear polariser, QWP: quarter-wave plate, HBT: Hanbury Brown-Twiss, SNSPDs: superconducting-nanowire single-photon detectors.  }
\end{figure*}
The exciton-phonon coupling is extracted from the strain profile after normalising the displacement by the zero-point motion. The QD couples to the strain via deformation potential
coupling~\cite{Munsch2017,Yeo2014}:   
\begin{equation}
    \Delta E = a(\epsilon_{xx}+\epsilon_{yy}+\epsilon_{zz})-\frac{b}{2}(\epsilon_{xx}+\epsilon_{yy}-2\epsilon_{zz}),
\end{equation}
where $\Delta E$ is the QD's energy shift, $a=-8.33~e\mathrm{V}$ and $b=-1.7~e\mathrm{V}$ are the deformation potentials for the hydrostatic and shear strain of GaAs, respectively~\cite{Vurgaftman2001,Walle1989}. The exciton-phonon coupling rate $g_\mathrm{ep}$ is: 
\begin{equation}
    g_\mathrm{ep} = \frac{\partial \omega_\mathrm{QD}}{\partial x}x_{\mathrm{zpf}}=\frac{\Delta E}{\hbar}.
    \label{eq_8}
\end{equation}
Figure~\ref{fig:1s}(e) shows $ g_\mathrm{ep}/2\pi$ for the in-plane breathing mode. In the very centre of the resonator, which is the optimal position for the QD location, we obtain $g_\mathrm{ep}/2\pi = $ \qty{3.2}{\MHz}. 

\section*{II. Device fabrication and measurement setup}
The wafer material is grown with molecular-beam epitaxy and consists of a \qty{1.15}{\um} AlAs sacrificial layer and a \qty{180}{\nm} GaAs diode structure. The diode consists of a QD-layer at the centre of the membrane (at $z=0$), as well as the p- and n-doped layers at the top and the bottom, respectively. The advantage of the QD-in-middle device is that the dots couple strongly to mechanical in-plane breathing modes. Above the QDs, there is an $\mathrm{Al}_\mathrm{0.33}\mathrm{Ga}_\mathrm{0.67}\mathrm{As}$ blocking layer, to minimise the diode leakage current. Details on the wafer material can be found in Ref.~\cite{spinnler_open_2023}.\\
\indent The mechanical resonator is fabricated by means of electron-beam lithography. First, the mesa structure is etched and $1.5\times$ \qty{1}{\um^2} contact pads are evaporated: Ni/Ge/Au/Ni/Au for the back contact (which is annealed to form an ohmic contact) and Cr/Au for the top contact. Second, using a soft mask, the nanostructures are written using an electron-beam and dry etched~\cite{Midolo2015, Uppu2020} into the membrane (inductively-coupled plasma reactive ion etching). Finally, after removing the residual resist, the structures are under-etched in a wet-etch process (hydrofluoric acid) and released via critical-point drying~\cite{Uppu2020}. During the fabrication of the mechanical resonators, the design axis (x-axis in Fig.~\ref{fig:1s}(d)) is aligned with the $[110]$ axis of the wafer. Since the mechanical properties of GaAs are anisotropic, the current mechanical mode would shift to \qty{1.25}{\GHz} when aligning to $[100]$.\\
\indent The sample is glued onto a titanium sample holder using non-conductive two-component epoxy (UHU, endfest 300) and the contacts are connected manually to a PCB using copper wires and silver epoxy (EPO-TEK, E4110). The sample is mounted on x/y/z-piezo steppers (attocube, ANPx101 \& ANPz101) and an x-y-scanner (attocube, ANSxy100lr) in a home-built vacuum-tube microscope with an optical NA = 0.65.
The tube is evacuated down to $4\times10^{-6}~\mathrm{mbar}$ and then filled with \qty{0.2}{mbar} helium exchange gas, corresponding to $2.8\times10^{-3}~\mathrm{mbar}$ at \qty{4.2}{\kelvin}. At this gas pressure, gas damping is negligible, see Ref.~\cite{spinnler_open_2023}. The measurement tube is precooled to liquid nitrogen temperature (\qty{77}{\kelvin}) and then moved into the He-bath cryostat (Cryovac) at \qty{4.2}{\kelvin}. The full measurement setup is shown in Fig.~\ref{fig:5s}.\\
\indent The gate voltage of the sample is controlled with a digital-to-analogue converter (Basel Precision Instruments, DAC SP 927). The optical excitation part of the setup consists of three different lasers: a diode laser for photoluminescence excitation at \qty{830}{\nm} (PicoQuant, LDH-D-C-830), a mode-locked laser for radiative-lifetime measurements (Coherent, Mira 900-D), and a tunable diode laser for resonant excitation around \qty{950}{\nm} (Toptica, DL pro). The resonant laser is frequency stabilised with a wavemeter (HighFinesse, WS7) and power stabilised with a double-pass acousto-optic modulator setup (Gooch and Housego, AOM 3200-1113 \& AODR 1200AF-AINA-2.5 HCR). All lasers are fibre-coupled (Thorlabs, SM-780HP) and sent to the cross-polarised optical microscope. The microscope consists of two polarising beam splitters, a linear polariser, and a quarter-waveplate. A laser suppression of up to $10^{-8}$ is typically achieved when the beam is focused on bulk GaAs~\cite{Kuhlmann2013a}. The sample surface can be imaged using a camera (Allied Vision, Guppy) in combination with an LED (Thorlabs, M940D2) and a removable pellicle beamsplitter (Thorlabs, BP145B2). The field of view is around \qty{10}{\mu m}. The collected QD photons are fibre-coupled and sent to either a spectrometer (Teledyne Princeton Instruments, Blaze 100HRX \& Acton SP2500i) or single-photon detectors in a Hanbury Brown-Twiss setup (Single Quantum, Eos \& Swabian Instruments, Time Tagger Ultra).

\section*{III. Optical characterisation of the quantum dot}
\begin{figure*}[ht]
    \centering
    \includegraphics[width=0.9\textwidth]{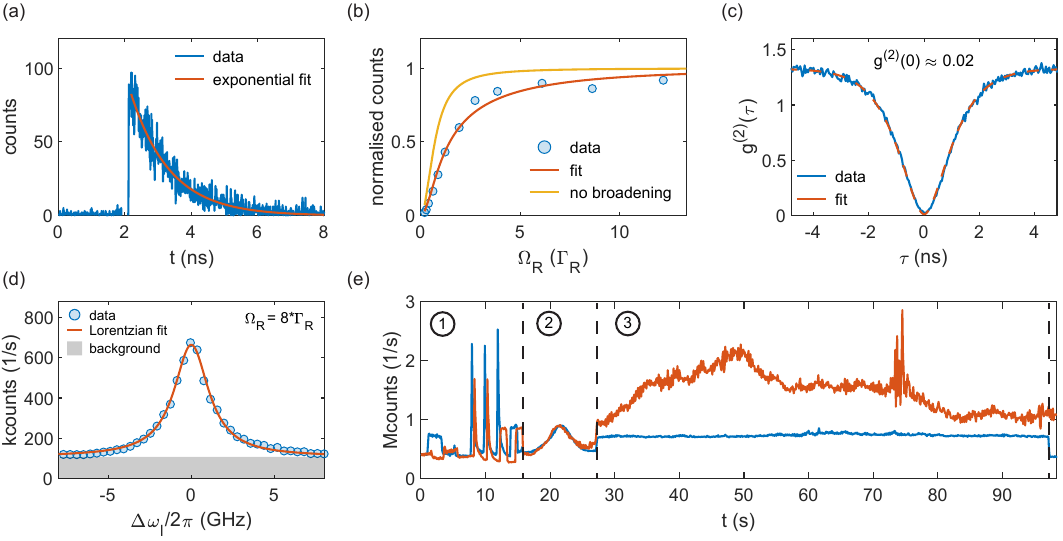}
\caption{\label{fig:2s} \textbf{Optical characterisation of the QD.} \textbf{(a)} Radiative-lifetime measurement with $\tau = $ \qty{1.18}{\ns}. The corresponding transform limit for the linewidth is $1/(2\pi\tau_\mathrm{R}) = $ \qty{135}{MHz}. \textbf{(b)} Resonant saturation-power curve. The excitation power is converted to the Rabi frequency, $\Omega_\mathrm{R}$, via a model fit to the data (orange curve). \textbf{(c)} Low-power ($\Omega_\mathrm{R}\ll\Gamma_\mathrm{R}$) autocorrelation measurement. Due to the high level of laser suppression, a single-photon purity of \qty{98}{\%} is achieved. \textbf{(d)} High-power resonant linewidth scan with $\Omega_\mathrm{R}=8~\Gamma_\mathrm{R}$. \textbf{(e)} Two example time traces (blue and red) of a high-power autocorrelation measurement: (1) automatic laser suppression, (2) locking of the QD resonance, and (3) autocorrelation measurement.   }
\end{figure*}

For the thermal-motion measurements, it is essential to find a QD that not only has a high exciton-phonon coupling rate but also a high count rate, a low inhomogeneous broadening, and good laser suppression. We present here additional QD characterisation measurements to those shown in the main text. In the following paragraphs (also in the main text), we refer to the frequency jittering of the QD resonance as the inhomogeneous broadening. The line broadening due to excited-state dephasing, we refer to as the homogeneous broadening of the QD, also see Ref.~\cite{Zhai2022}. \\
\indent To estimate the inhomogeneous broadening, we compare the measured linewidth to the transform limit, see Fig.~2(d) in the main text. To determine the lifetime, we excite the QD using picosecond optical pulses. The time tagger module is synchronised with the pulsed laser and a time histogram is recorded, see Fig.~\ref{fig:2s}(a). The excited state of the QD ($X^{1-}$) freely decays with a time constant of $\tau_\mathrm{R} = $ \qty{1.18}{\ns}. This gives an excited-state decay rate of $\Gamma_\mathrm{R}= 1/\tau_\mathrm{R} = $ \qty{847}{\MHz}, with a corresponding transform-limited linewidth of $\Gamma_\mathrm{R}/2\pi = $ \qty{135}{\MHz}. A low-power linewidth measurement is presented in the main text, which yields $\Gamma_\mathrm{inh}/2\pi = $ \qty{550}{\MHz}, which is a factor of four above the transform limit.\\
\indent The Rabi frequency, $\Omega_\mathrm{R}$, describes the interaction strength of the laser with the QD. We convert the excitation power to the Rabi frequency by carrying out a resonant power-saturation measurement, see Fig.~\ref{fig:2s}(b). Each data point represents the peak intensity of a linewidth scan (obtained from a Lorentzian fit) at the corresponding excitation power. The QD count rate is proportional to the excited-state population, $\rho_\mathrm{ee}$, which is given by~\cite{spinnler_open_2023}: 
\begin{equation}
\langle \hat{\sigma}_\mathrm{+} \hat{\sigma}_\mathrm{-}\rangle=  \rho_{ee} = \frac{\left(\frac{1}{2}\Omega_\mathrm{R}\right)^2}{\Delta\omega_\mathrm{l}^2+\frac{1}{2}\Omega_\mathrm{R}^2+\left(\frac{1}{2}\Gamma_\mathrm{R}\right)^2},
\label{eq:rho_ee}
\end{equation}
where $\Omega_\mathrm{R}$ is the Rabi frequency, $\Delta\omega_\mathrm{l}/2\pi$ is the laser detuning from the QD transition, and $\Gamma_\mathrm{R}$ is the excited-state decay rate. By fitting Eq.~\ref{eq:rho_ee} to the data, the excitation power is translated to $\Omega_\mathrm{R}$. Here, we include the inhomogeneous broadening by a convolution of Eq.~\ref{eq:rho_ee} with a Lorentzian-weighted detuning jitter of \qty{400}{\MHz}. Without this, the Rabi frequency would be underestimated, yellow line in Fig.~\ref{fig:2s}(b).\\
\indent The ratio between unsuppressed laser and QD counts depends highly on the laser spot position which in turn depends on the QD position. Furthermore, it also highly depends on the excitation power. Figure~\ref{fig:2s}(c) shows a low-power autocorrelation measurement, with \qty{5}{\nW} of laser power reaching the sample ($\Omega_\mathrm{R}\ll\Gamma_\mathrm{R}$). The high single-photon purity of \qty{98}{\%} proves that the QD acts as a single-photon emitter. The autocorrelation is fitted with the standard autocorrelation function of a two-level system~\cite{Jahn2015}.\\
\indent At high excitation powers ($\Omega_\mathrm{R}\gg\Gamma_\mathrm{R}$) the emission of the quantum emitter saturates, however, the unsuppressed laser increases. This leads to a reduced signal-to-background level. Figure~\ref{fig:2s}(d) shows a high-power linewidth scan, where the background level is significant (compared to Fig.~2(d) in the main text). The signal-to-background increases further upon detuning the laser from resonance. Therefore, in an autocorrelation measurement, the higher the laser power and the higher the laser detuning, the lower the single-photon purity (higher $g^{(2)}(0)$). In addition, the background level is also very unstable. Figure~\ref{fig:2s}(e) shows two example time traces of a high-power ($\Omega_\mathrm{R} = 8~\Gamma_\mathrm{R}$) autocorrelation measurement with a laser detuning of $\Delta\omega_\mathrm{l}/2\pi = $ \qty{1}{\GHz}. The measurement is performed as follows. First, the laser is automatically suppressed by alternately optimising the linear polariser and quarter-wave-plate angles in our dark-field microscope. Second, we perform a linewidth scan and lock the QD resonance to the laser frequency, thus, compensating for spectral drifts. Third, autocorrelation data is recorded for one minute, followed by going back to step one. During the measurement, the laser suppression can drift up to an order of magnitude, a result of a slight change in laser spot position due to vibrational noise from the environment and pressure changes in the
helium recovery line.

\section*{IV. From autocorrelation to noise-power spectrum}
\begin{figure*}[ht]
    \centering
    \includegraphics[width=0.9\textwidth]{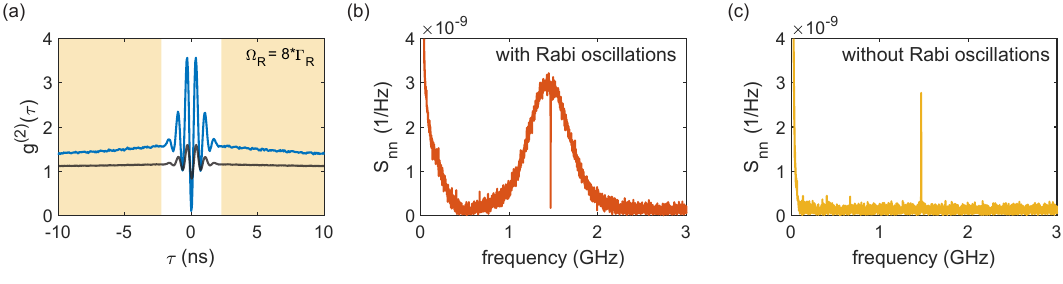}
\caption{\label{fig:3s} \textbf{From autocorrelation to noise-power spectrum.} \textbf{(a)} Autocorrelation measurement at optimal detuning, $\Omega_\mathrm{R}^\mathrm{eff} = \Omega_\mathrm{m}$, before (black) and after (blue) correcting for the unsuppressed laser. This is the same measurement as in the main text in Fig.~3(h). \textbf{(b)} Fourier transform of the full autocorrelation data which is shown in (a). \textbf{(c)} Fourier transform of the autocorrelation data without Rabi oscillations, yellow shaded area in (a).   }
\end{figure*}

In our experiments, we acquire the noise-power spectrum via an autocorrelation measurement. As mentioned above, the collected signal contains unsuppressed laser light, which does not carry information about the mechanical resonator. Therefore, to obtain the true mechanical noise power, we need to correct for the unsuppressed laser. The unsuppressed laser results in a flat background in the autocorrelation measurement (see Fig.~\ref{fig:3s}(a), black curve), which we correspondingly subtract. Subsequently, the autocorrelation is normalised to one at large time delays (ms-regime). Figure~\ref{fig:3s}(a) shows the post-processed autocorrelation, performed at optimal detuning such that $\Omega_\mathrm{R}^\mathrm{eff} = \Omega_\mathrm{m}$. At long time delays, weak oscillations due to the interaction with the mechanical resonator are visible (also see main paper). 
The prominent oscillations at short delays are Rabi oscillations at $\Omega_\mathrm{R}^\mathrm{eff}$. The noise-power spectrum is related to the autocorrelation via a Fourier transform (Wiener–Khinchin theorem)~\cite{Munsch2017}: 
\begin{equation}
    S_{\mathrm{nn}}(f) = 2\mathrm{FFT}\left[g^{(2)}(\tau)\right]\tau_{\mathrm{bin}},
    \label{eq:FFT}
\end{equation}
where $g^{(2)}(\tau)$ is the normalised autocorrelation data, and $\tau_\mathrm{bin}$ is the autocorrelation binning time. Figure~\ref{fig:3s}(b) shows $S_{\mathrm{nn}}(f)$ obtained from the full autocorrelation data. The broad peak around \qty{1.5}{\GHz} is associated with the strong but rapidly decaying Rabi oscillations from Fig.~\ref{fig:3s}(a). The sharp feature is due to the QD-mechanical interaction. Due to a phase difference of $\pi$ between Rabi oscillations and mechanical modulation, the mechanical-noise peak appears as a dip in the broad Rabi peak. This makes it hard to integrate the mechanical-noise peak. To remove the Rabi oscillations, we perform the Fourier transform on the data at delays $\tau>\tau_\mathrm{R}$ (yellow region in Fig.~\ref{fig:3s}(a)), once the Rabi oscillations are completely damped. The corresponding noise-power spectrum shows the mechanical noise as a peak on a flat background, see Figure~\ref{fig:3s}(c). This process does not compromise the mechanical noise power since the mechanical damping constant of $\tau_\mathrm{m} = 2Q_\mathrm{m}/ \Omega_\mathrm{m} = $ \qty{0.46}{\us} is several orders of magnitudes larger than the Rabi damping of $\approx$ \qty{1}{\ns}.

\section*{V. Master-equation simulations}
To simulate the exciton-phonon interaction we perform master-equation simulations. The full Hamiltonian of the system is given by: 
\begin{equation}
\hat{H} = \hat{H}_\mathrm{QD} + \hat{H}_\mathrm{m} + \hat{H}_\mathrm{int} + \hat{H}_\mathrm{drive},
\end{equation}
where $\hat{H}_\mathrm{QD}$ is the QD, $\hat{H}_\mathrm{m}$ the mechanical, $\hat{H}_\mathrm{int}$ the interaction, and $\hat{H}_\mathrm{drive}$ the optical drive part. We describe the QD as a simple two-level system (TLS) with a ground and an excited state, $\ket{g}$ and $\ket{e}$, respectively. The TLS is driven by a classical optical field. In the dipole approximation, this reads: 
\begin{equation}
\hat{H}_\mathrm{QD} + \hat{H}_\mathrm{drive} = \hbar\omega_\mathrm{QD}\hat{\sigma}_\mathrm{+}\hat{\sigma}_\mathrm{-} -\hbar\frac{\Omega_\mathrm{R}}{2}\left( \hat{\sigma}_\mathrm{+}+\hat{\sigma}_\mathrm{-}\right) \left( e^{i\omega_\mathrm{l}t}+e^{-i\omega_\mathrm{l}t}\right),
\end{equation}
where $\hbar\omega_\mathrm{QD}$ is the QD's exciton transition energy, $\Omega_\mathrm{R}$ is the optical Rabi frequency, $\omega_\mathrm{l}/2\pi$ is the frequency of the driving field, $\hat{\sigma}_\mathrm{+} = \ket{e}\bra{g}$ and $\hat{\sigma}_\mathrm{-} = \ket{g}\bra{e}$ are the Pauli transition operators. The mechanical part is described by a quantum harmonic oscillator: 
\begin{figure*}[ht]
    \centering
    \includegraphics[width=0.9\textwidth]{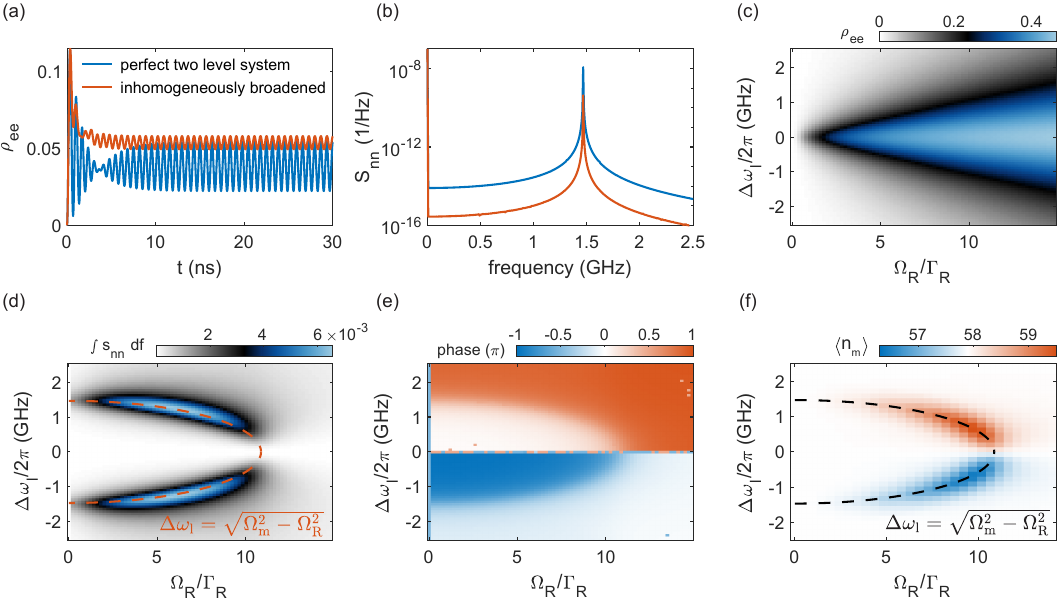}
\caption{\label{fig:4s} \textbf{Master equation simulations} with a classical (a-e) and a quantum description (f) of the mechanical resonator. \textbf{(a)} Numerical simulation of the excited-state population, showing a modulation due to the mechanical vibrations. \textbf{(b)} Fourier transform of the data shown in (a). \textbf{(c)} Steady-state (time-averaged) excited-state population in dependence of the laser detuning and Rabi frequency. \textbf{(d)} Integrated mechanical-noise peak, and \textbf{(e)} phase of the modulation. \textbf{(f)} Expectation value of the phonon occupation upon detuning the laser blue (heating) or red (cooling). The simulation parameters are: $\tau_\mathrm{R} = $ \qty{1.18}{\ns}, $\Omega_\mathrm{m}/2\pi = $ \qty{1.466}{\GHz}, $g_\mathrm{ep}^\mathrm{th}/2\pi $ = \qty{34.4}{\MHz}, $g_\mathrm{ep}/2\pi $ = \qty{3.2}{\MHz}, and $\langle n_\mathrm{m}\rangle = 58$. }
\end{figure*}
\begin{equation}
    \hat{H}_\mathrm{m} = \hbar\Omega_\mathrm{m}\left( \hat{b}^\dagger  \hat{b} +1/2\right),
\end{equation}
where $\Omega_\mathrm{m}/2\pi$ is the mechanical frequency, $\hat{b}^\dagger$ and $\hat{b}$ are the phonon creation and annihilation operators, respectively. The phonon occupation of the mechanical resonator is $\langle n_\mathrm{m}\rangle = \langle\hat{b}^\dag\hat{b}\rangle$. The interaction part between the two systems has a dispersive form where the displacement leads to a shift in the excited-state energy of the QD:
\begin{equation}
    \hat{H}_\mathrm{int} = \hbar g_\mathrm{ep}\hat{\sigma}_\mathrm{+}\hat{\sigma}_\mathrm{-}\left( \hat{b}^\dagger  + \hat{b} \right),
\end{equation}
where $g_\mathrm{ep}/2\pi$ is the exciton-phonon coupling rate, $\hat{\sigma}_\mathrm{+}\hat{\sigma}_\mathrm{-} = \ket{e}\bra{e}$, and $(\hat{b}^\dagger+\hat{b}) = \hat{x}/x_\mathrm{zpf}$ is the displacement operator in units of the zero-point motion. The interaction part can also be described classically: 
\begin{equation}
    \hat{H}_\mathrm{int} = \hbar \frac{g_\mathrm{ep}}{x_\mathrm{zpf}}x_\mathrm{th}\sin{\left(\Omega_\mathrm{m}t\right)}\hat{\sigma}_\mathrm{+}\hat{\sigma}_\mathrm{-}= \hbar g_\mathrm{ep}^\mathrm{th}\sin{\left(\Omega_\mathrm{m}t\right)}\hat{\sigma}_\mathrm{+}\hat{\sigma}_\mathrm{-} ,
\end{equation}
where $x_\mathrm{th}$ is the thermal displacement, and $g_\mathrm{ep}^\mathrm{th}=g_\mathrm{ep}\sqrt{2\langle n_\mathrm{m}\rangle}=2\pi\times $\qty{34.4}{\MHz} is the thermal exciton-phonon coupling rate~\cite{Leijssen2017}. In the rotating frame of the laser field, the full Hamiltonian reads: 
\begin{equation}
    \hat{H}_\mathrm{RWA} = -\hbar\Delta\omega_\mathrm{l}\hat{\sigma}_\mathrm{+}\hat{\sigma}_\mathrm{-}+\hbar\Omega_\mathrm{m}\left(\hat{b}^\dagger \hat{b}+1/2\right)+\hbar g_\mathrm{ep}\hat{\sigma}_\mathrm{+}\hat{\sigma}_\mathrm{-}\left(\hat{b}^\dagger+\hat{b}\right) +\hbar\frac{\Omega_\mathrm{R}}{2}\left(\hat{\sigma}_\mathrm{+}+\hat{\sigma}_\mathrm{-} \right),
    \label{eq:RWA}
\end{equation}
where $\hbar\Delta\omega_\mathrm{l}$ is the energy detuning between the driving field and the QD transition. The incoherent part of the Hamiltonian, which is the QD's radiative decay, is added via a Lindblad operator $\hat{L}= \sqrt{\Gamma_\mathrm{R}}\hat{\sigma}_\mathrm{-}$. The time dynamic of the system is captured by the von Neumann equation~\cite{Gerry2004}:
\begin{equation}
\frac{\partial}{\partial t}\hat{\rho} = -\frac{i}{\hbar}[ \hat{H}_\mathrm{RWA} ,\hat{\rho}] + \hat{\mathcal{L}}(\hat{\rho}),\;\;\;\;\hat{\mathcal{L}}(\hat{\rho}) = \frac{1}{2}\left(2\hat{L}\hat{\rho}\hat{L}^\dag-\hat{\rho}\hat{L}^\dag \hat{L}-\hat{L}^\dag \hat{L}\hat{\rho} \right).
    \label{eq:vN}
\end{equation}
\indent In our first simulation, we reproduce the time-modulation in the QD's emission upon detuning the probe laser field. Since the phonon population is large, $\langle n_\mathrm{m}\rangle = 58\gg1$, the interaction can be expressed classically~\cite{Weiss2021}, excluding backaction on the mechanical resonator. Furthermore, we assume that the phase of the mechanical resonator is static on the time scales of the QD dynamics (few ns).  We perform numerical simulations using Eq.\ref{eq:RWA}-\ref{eq:vN} and solve for the excited-state population, $\rho_\mathrm{ee}$. The inhomogeneous broadening is included by a Lorentzian-weighted detuning jitter of \qty{400}{\MHz}. Figure~\ref{fig:4s}(a) shows $\rho_\mathrm{ee}$ as a function of time without (blue) and with (orange) inhomogeneous broadening, respectively. Laser detuning and Rabi frequency are chosen such that $\Omega_\mathrm{R}^\mathrm{eff} = \Omega_\mathrm{m}$. The strong oscillations at short $t$ are the optical Rabi oscillations and the weaker oscillations arise due to the QD-mechanical coupling. Note that, in general, the mechanical modulation shows a higher amplitude in the time trace than in the autocorrelation. Around \qty{3}{\ns}, the phase shifts from the Rabi to the mechanical oscillations. The simulation is performed over $300$ mechanical periods, where we analyse further only the final $50$ periods. As for the autocorrelation, we obtain the noise-power spectrum via a Fourier transform of the normalised time trace: 
\begin{equation}
    S_\mathrm{nn}(f) = 2\mathrm{FFT}\left[\frac{\rho_\mathrm{ee}(t)}{\langle \rho_\mathrm{ee} \rangle}\right]^2\frac{t_{\mathrm{bin}}^2}{t_{\mathrm{sim}}},
    \label{eq:FFT_tt}
\end{equation}
where $\langle \rho_\mathrm{ee} \rangle$ is the average excited-state population, $t_\mathrm{bin}$ is the binning time, and $t_\mathrm{sim}$ is the length of the simulation. Figure~\ref{fig:4s}(b) shows $S_\mathrm{nn}(f)$ obtained from the simulation shown in Fig.~\ref{fig:4s}(a). Comparing the simulations without and with inhomogeneous broadening, we observe an order-of-magnitude reduction in signal strength. \\
\indent The numerical simulations are performed upon sweeping the Rabi frequency and laser detuning. Figure~\ref{fig:4s}(c) shows the time-averaged excited-state population. The integrated noise-power of the mechanical modulation is shown in Fig.~\ref{fig:4s}(d). As can be seen, the highest interaction between the two systems is found when the effective Rabi frequency matches the mechanical frequency: 
\begin{equation}
    \Omega_\mathrm{R}^\mathrm{eff} = \sqrt{\Omega_\mathrm{m}^2+\Delta\omega_\mathrm{l}^2} \mbeq \Omega_\mathrm{m}.
\end{equation}
At low excitation powers ($\Omega_\mathrm{R}\ll\Omega_\mathrm{m}$), this corresponds to detuning the laser to one of the acoustic sidebands $\Delta\omega_\mathrm{l}=\pm\Omega_\mathrm{m}$. Conversely, at high excitation powers ($\Omega_\mathrm{R}>\Omega_\mathrm{m}$), this corresponds to $\Delta\omega_\mathrm{l} = \sqrt{\Omega_\mathrm{m}^2-\Omega_\mathrm{R}^2}$ (highlighted in orange). Figure~\ref{fig:4s}(e) shows the phase of the excited state's time modulation. Several transitions of $\pi$ are visible, matching with the resonance condition of $\Omega_\mathrm{R}^\mathrm{eff} = \Omega_\mathrm{m}$. Note that the signal-to-noise ratio in the measured power spectrum depends on the product of Fig.~\ref{fig:4s}(c) and (d), which is the noise sensitivity multiplied by the excited-state population. \\
\indent In order to extract the degree of mechanical cooling, we perform a second simulation where we include the backaction on the phonon population of the mechanical resonator. For this, the mechanical resonator is treated as a quantum harmonic oscillator coupled to a thermal bath. To reduce simulation time, we make use of the quantum simulation toolbox Qutip~\cite{Johansson2012,Johansson2013}. We perform master-equation simulations using Eq.~\ref{eq:RWA} and \ref{eq:vN}, solving for the steady-state solution. The coupling to the thermal bath is described with additional collapse operators:   
\begin{equation}
    \hat{c}_\mathrm{in} = \sqrt{\Gamma_\mathrm{m}\cdot n_\mathrm{bath}}\cdot\hat{b}^\dagger,\;\;\;\;\;\,\hat{c}_\mathrm{out} = \sqrt{\Gamma_\mathrm{m}\cdot\left(n_\mathrm{bath}+1\right)}\cdot\hat{b},
\end{equation}
where $\Gamma_\mathrm{m} = \Omega_\mathrm{m}/Q_\mathrm{m}$ is the energy dissipation rate, and $n_\mathrm{bath} = 58$ is the thermal-bath population (at $\Omega_\mathrm{m}$). The simulations are performed with an $N_\mathrm{m}=500$ dimensional mechanical Hilbert space. Also here, \qty{400}{\MHz} of inhomogeneous broadening is included. Figure~\ref{fig:4s}(f) shows the expectation value of the resonator's phonon population, $\langle n_\mathrm{m}\rangle$, as a function of Rabi frequency and laser detuning. As before, the strongest interaction is found when $\Omega_\mathrm{R}^\mathrm{eff}=\Omega_\mathrm{m}$. To observe a change in phonon population, high excitation powers and laser detunings smaller than $\Omega_\mathrm{m}$ are required. With the current exciton-phonon coupling rate, the change in phonon number is small. However, this can be changed either by increasing $g_\mathrm{ep}$ or $Q_\mathrm{m}$, or by reducing $\Gamma_\mathrm{inh}$, see discussion in the main text.

\end{document}